\documentclass{emulateapj}
\newcommand{\etal}{{et al.\ }}
\newcommand{\Kepler}{{\sl Kepler}\ }
\newcommand{\be}{\begin{equation}}
\newcommand{\ee}{\end{equation}}

\slugcomment{accepted to ApJ}
\shorttitle{TTV Candidates from \Kepler}
\shortauthors{Ford et al.}

\begin{document}
\title{
Transit Timing Observations from \Kepler:
V. Transit Timing Variation Candidates in the First Sixteen Months from Polynomial Models
}


\author{
Eric B. Ford\altaffilmark{1},        
Darin Ragozzine\altaffilmark{2},    
Jason F. Rowe\altaffilmark{3,4},    
Jason H. Steffen\altaffilmark{5},   
Thomas Barclay\altaffilmark{3,6},
Natalie M. Batalha\altaffilmark{7}, 
William J. Borucki\altaffilmark{3}, 
Stephen T. Bryson\altaffilmark{3},  
Douglas A. Caldwell\altaffilmark{3,4},
Daniel C. Fabrycky\altaffilmark{8,9}, 
Thomas N. Gautier III\altaffilmark{10},
Matthew J. Holman\altaffilmark{2},  
Khadeejah A. Ibrahim\altaffilmark{11},
Hans Kjeldsen\altaffilmark{12},
Karen Kinemuchi\altaffilmark{3,6},
David G. Koch\altaffilmark{3}, 
Jack J. Lissauer\altaffilmark{3},   
Martin Still\altaffilmark{3,6},
Peter Tenenbaum\altaffilmark{3,4},
Kamal Uddin\altaffilmark{11},
William Welsh\altaffilmark{13}      
}
\altaffiltext{1}{Astronomy Department, University of Florida, 211 Bryant Space Sciences Center, Gainesville, FL 32111, USA}
\altaffiltext{2}{Harvard-Smithsonian Center for Astrophysics, 60 Garden Street, Cambridge, MA 02138, USA}
\altaffiltext{3}{NASA Ames Research Center, Moffett Field, CA, 94035, USA}
\altaffiltext{4}{SETI Institute, Mountain View, CA, 94043, USA}
\altaffiltext{5}{Fermilab Center for Particle Astrophysics, P.O. Box 500, MS 127, Batavia, IL 60510}
\altaffiltext{6}{Bay Area Environmental Research Institute, 560 Third St West, Sonoma, CA 95476, USA}
\altaffiltext{7}{San Jose State University, San Jose, CA 95192, USA}
\altaffiltext{8}{UCO/Lick Observatory, University of California, Santa Cruz, CA 95064, USA}
\altaffiltext{9}{Hubble Fellow}
\altaffiltext{10}{Jet Propulsion Laboratory/California Institute of Technology, Pasadena, CA 91109, USA}
\altaffiltext{11}{Orbital Sciences Corporation/NASA Ames Research Center, Moffett Field, CA 94035, USA}
\altaffiltext{12}{Department of Physics and Astronomy, Aarhus University, DK-8000 Aarhus C, Denmark}
\altaffiltext{13}{San Diego State University, 5500 Campanile Drive, San Diego, CA 92182-1221, USA}
\email{eford@astro.ufl.edu}

\begin{abstract}
Transit timing variations provide a powerful tool for confirming and characterizing transiting planets, as well as detecting non-transiting planets.  
We report the results an updated TTV analysis for 1481 planet candidates (Borucki \etal 2011; Batalha \etal 2012) based on transit times measured during the first sixteen months of \Kepler observations.  We present 39 strong TTV candidates based on long-term trends (2.8\% of suitable data sets).  We present another 136 weaker TTV candidates (9.8\% of suitable data sets) based on excess scatter of TTV measurements about a linear ephemeris.  We anticipate that several of these planet candidates could be confirmed and perhaps characterized with more detailed TTV analyses using publicly available \Kepler observations.  For many others, \Kepler has observed a long-term TTV trend, but an extended \Kepler mission will be required to characterize the system via TTVs.  We find that the occurrence rate of planet candidates that show TTVs is significantly increased ($\sim$68\%) for planet candidates transiting stars with multiple transiting planet candidate when compared to planet candidates transiting stars with a single transiting planet candidate.   
\end{abstract}


\section{Introduction} 
\label{secIntro}
Long-term, high-precision photometric observations from NASA's \Kepler mission provide the basis for confirming and characterizing planets via transit timing variations (TTVs).  
\Kepler has confirmed 10 transiting planets via TTVs (Holman \etal 2010; Ballard \etal 2011; Cochran \etal 2011; Lissauer \etal 2011a).  
Many more \Kepler planet candidates are expected to reveal significant TTVs, as the time baseline of \Kepler observations grows (Ford \etal 2011; hereafter F11).  
F11 reported preliminary indications of TTVs for dozens of additional planet candidates and predicted significant TTVs for several planet candidates in multiple transiting planet candidate systems based on the first fourth months of \Kepler observations.  
This paper and Steffen \etal (2012b) provide updated TTV analyses of \Kepler planet candidates to identify planets candidates worthy of more detailed TTV analyses.  
We describe our methods in \S\ref{secMethods}.  Table \ref{tabKey} provides a description of the online-only, machine-readable material.  We provide an overview of our results and discuss a subset of particularly interesting systems in \S\ref{secDiscuss}.

\section{Methods}
\label{secMethods}
We analyze transit times observed by \Kepler through the end of quarter six (September 20, 2010; J. Rowe \etal in preparation; hereafter R12).  
We include all active planet candidates from Borucki \etal (2011) and Batalha \etal (2012) for which there are at least three transits and the photometric signal-to-noise of a typical transit is at least three. 

For each planet candidate to be analyzed, we calculate the best-fit linear ephemeris (L$_1$) to the transit times and uncertainties from R12.  We anticipate that the distribution of transit timing errors is likely to approximated by a mixture distribution with heavily weighted Gaussian component (core) and a weakly weighted distribution with broader tails, due to the difficulties of measuring transit times (e.g., data artifacts, stellar variability, nearly simultaneous transit of another planet, large transit timing variations, or other complications).  In order to reduce the effects of a small number of poorly measured transit times, we do not report the initial best-fit linear ephemeris (L$_1$), but rather report an updated linear ephemeris (L$_2$) calculated after cleaning the transit times to exclude outliers and transit times with unusually large uncertainties.  Specifically, we reject any transit times with an absolute deviation from the linear ephemeris exceeding four times the median absolute deviation of transit times from the first linear ephemeris (L$_1$).  We also reject any transit times for which the measurement uncertainty exceeds twice the median of the published transit time uncertainties.  We record the median timing uncertainty for the remaining $N_{TT}$ transit times ($\sigma_{TT}$).  Then, we calculated an updated the best-fit linear ephemeris (L$_2$) using the cleaned transit times.  We report the epoch ($E_{\mathrm{lin}}$) and period ($P_{\mathrm{lin}}$) of the updated linear ephemeris (L$_2$) in the electronic table.
 
First, we identify planet candidates with larger than expected scatter about the updated linear ephemeris (L$_2$) in the measured transit times based on the measurement uncertainties (hereafter ``excess scatter'').  We calculate the median absolute deviation (MAD) and the weighted root mean square deviation (WRMS) of the clipped transit times relative to the updated linear ephemeris (F11).  We calculate the $p$-value for a standard $\chi^2$ test, $p_{\chi^2,\mathrm{lin}}$ which is the probability that we would measure a $\chi^2$ as large or larger than $\chi^2_{\mathrm{lin}}$ assuming transit timing errors follow a Gaussian distribution.   We also calculate ($p_{\chi'^2,\mathrm{lin}}$), the p-value for a $\chi^2$-test using $\chi'^{2}=\pi~N_{TT} (\mathrm{MAD})^2/(2\sigma_{TT}^2)$, as described in F11.  Since we cannot be confident that the transit time errors follow a Gaussian distribution, one should interpret the listed $p$-values with caution (particularly for low S/N transits).  We consider a small $p$-value as indicating that a transit timing data set is worthy of further investigation.  We identify planet candidates with a $p_{\chi^2,\mathrm{lin}}<10^{-3}$ and/or $p_{\chi'^2,\mathrm{lin}}<10^{-3}$ by setting the 1 and 2 bits of the TTV flag (see electronic table).  For a few planet candidates, the transit times have a small number of transits, but each has a highly significant deviation from a linear ephemeris.  We indicate these (MAD$\ge~5\sigma_{TT}$) with the 3rd bit (with a value of 4) of the TTV flag.

Second, we identify planet candidates with long-term trends in the deviations of the measured transit times from a linear ephemeris (hereafter referred to as ``long-term TTV trends'').  We calculate the best-fit (minimum $\chi^2$) $n_d$-degree polynomial ephemeris
\be
\hat{t}_n = E_{n_d} + n \times P_{n_d} \left(1 + \sum_{i=2}^{n_d} c_i n^{i-1} \right),
\ee
where $\hat{t}_n$ is the predicted transit time of the $n$th transit, $E_{n_d}$ and $P_{n_d}$ are the epoch and period of the $n_d$-degree model and the $c_i$'s are the higher-order coefficients.  Note that we renumber the transits relative to R12, so the epoch falls near the middle of the observed transits, reducing covariance between the epoch and period.  
We calculate $\chi^2_{n_d}$ for each $n_d$ up to 4, if there are at least 8 transits and otherwise for $n_d<$nTT-2.  We perform standard $F$-tests using the ratio of $\chi^2$ for the $n_d$-degree polynomial model and all lower-degree ephemerides.  If $p_{F,\mathrm{quad,lin}}$, the $p$-value for the $F$-test comparing the linear and quadratic models is less than 2\%, then we conclude that the difference in $\chi^2_{\mathrm{quad}}$ and $\chi^2_{\mathrm{lin}}$ is greater than one would expect due to mere fluctuations at the 2\% significance level.  Therefore, we accept the quadratic model and set the 4th bit (with a value of 8) of the TTV Flag in the electronic table .  We repeat the procedure for the cubic and then the quartic models, accepting the higher degree models if the $F$-test $p$-value is less than 2\% when comparing the higher degree model to both the linear model and the last accepted model.  We list $N_d$, the degree of the highest-degree accepted model, along with $p$-values for the various tests in the electronic table.

\section{Results}
\label{secResults}
%
We identified 39 TTV candidates with long-term trends using simple polynomial models (see Figures 1, 2 and 3)\footnote{Similar figures for additional planet candidates will be available at \hbox{http://www.astro.ufl.edu/$\sim$eford/data/kepler/}.}.  Of these, we approximate the TTVs with 
a quadratic model for 14 planet candidates (KOIs 157.04, 168.01, 227.01, 308.01, 448.02, 473.01, 784.01, 886.01, 1081.01, 1529.01, 1573.01, 1581.01, 1599.01, 1840.01), 
%
%
a cubic model for 14 planet candidates 
(KOIs 456.01, 524.01, 738.01, 806.03, 854.01, 884.02, 918.01, 1102.02, 1145.01, 1199.01, 1270.02, 1366.02) 
%
%
and a quartic model for 11 planet candidates (KOIs 103.01, 137.02, 142.01, 244.02, 248.01, 277.01, 377.01, 377.02, 841.01, 984.01, 1102.01).  
%
%
Already, KOIs 137.02 (Kepler-18d), 157.04 (Kepler-11f), 168.01 (Kepler-23c), 244.02 (Kepler-25b), 248.01, 377.01 (Kepler-9b), 377.02 (Kepler-9c), 738.01 (Kepler-29b), 806.03 (Kepler-30b), 841.01 (Kepler-27b), 1102.01 (Kepler-24c), 1102.02 (Kepler-24b) have been confirmed (Cochran et al.\ 2011; Lissauer et al.\ 2011a; Ford et al.\ 2012; Steffen et al.\ 2012; Holman et al.\ 2010; Fabrycky et al.\ 2012), most occurring between this paper's original submission and publication.  

Five planet candidates have extremely significant TTVs, but polynomial models did not provide a significant improvement in fit, due to the limited number of transits (351.02, 806.01, 1271.01 and 1474.01) or a complex signal that was not well approximated by a $\le~4$th degree polynomial (872.01).  While we regard these as relatively strong TTV candidates, more detailed analyses will be necessary before most of the planets are confirmed, as one must ensure that the TTVs are dynamical in origin and that we are not observing eclipse timing variations in a triple star system (Carter \etal 2011; Slawson \etal 2011; Steffen \etal 2011) that has been diluted so the eclipse depth is consistent with a planet transit.  Already, KOIs 806.01 (Kepler-30d) and 872.01 (KOI-872b) have been promptly confirmed between this paper's original submission and publication (Fabrycky et al.\ 2012; Nesvorny et al.\ 2012).

For an additional 131 planet candidates, the measured transit times appear to have a scatter greater than that expected based on the measurement uncertainties.  While we regard these as weaker TTV candidates, we are optimistic that many may be closely-packed multiple planet systems for which TTVs could characterize the planet masses and orbital properties.  For example, the electronic table identifies several confirmed planets as having excess TTV scatter but without a significant long-term trend (KOI 72.01=Kepler-10b, KOI 137.01=Kepler-18c, KOI 157.03=Kepler-11e, KOI 203.01=Kepler-17b, KOI 244.01=Kepler-25c, KOI 250.01=Kepler-26b, KOI 806.02=Kepler-30c, KOI 952.02=Kepler-32c).  
Thus, we recommend these planet candidates with excess TTV scatter for further analysis, as well as long-term monitoring by \Kepler using the 1-minute cadence to improve the number, accuracy and precision of transit time measurements.  

We note that more detailed analyses have detected TTVs for several planets that were not identified as having a long-term trend or excess scatter by our analysis (e.g., Kepler-11b, Kepler-19b).  Thus, we anticipate that more detailed analyses (particularly those that use short-cadence observations or systems with multiple transiting planet candidates) may confirm and characterize additional planets via TTVs.

This analysis significantly improves on that of F11, primarily due to the increased number and time span of transit timing observations now available.  
Of those with an obvious TTV pattern (TTV Flag of 1 in F11 Table 5), five of seven are identified as significant in our updated analysis.  One exception (KOI 928.01) was not included in our analysis, as it is now a known triple star (Steffen \etal 2011b).  For the other exception (KOI 528.01), we now rule out a long-term TTV trend, but low amplitude quaesiperidic TTVs are still a possibility.  The remaining five TTV candidates have a statistically significant long-term trend once we include observations through Q6.  Of the seven candidates with $p_{F,\mathrm{quad,lin}}\le0.03$ from F11 (Table 4), only two (KOI 142.01, 227.01) were identified as TTV candidates in our updated analysis.  As noted in F11, the small number of transits and small time span of observations was a significant limitation.

\section{Discussion}
\label{secDiscuss}
%
%
We compare the frequency of TTV candidates according to orbital period, transit S/N and whether the host star has multiple transiting planet candidates.  For this purpose, we restrict our attention to planet candidates with at least 5 transits (after clipping points with unusually large uncertainties or absolute deviations, as described in \S\ref{secMethods}), leaving 1386 TTV data sets, including 851 (535) planet candidates associated with KOIs that have a single (multiple) transiting planet candidates. 
%
%
We find 163 (18\%) TTV data sets with some indication of potential TTVs (TTV Flag$\ge$1), including 89 (10.5\%) and 74 (13.8\%) for stars with single and multiple transiting planet candidates, respectively.  
If we restrict our attention on systems with more robust signature of TTVs (TTV Flag$\ge$2), then we find 109 (7.9\%) TTV data sets had excessive scatter including 57 (6.7\%) and 52 (9.7\%) for single and multiple candidate systems, respectively.  Based on Monte Carlo simulations, the probability of these two subsets of KOIs have the same occurrence rate of long term TTV trends is 0.7\%.

%
%
We find 39 (2.8\%) TTV data sets with a significant long-term trend ($N_d>=2$), including 19 (2.2\%) and 20 (3.7\%) for stars with single and multiple transiting planet candidates, suggesting a $\sim$68\% greater rate of long-term TTV trends in systems with multiple transiting planet candidates. Based on Monte Carlo simulations, the probability of these two subsets of KOIs have the same occurrence rate of long term TTV trends is 2.6\%.  Therefore, there is significant evidence that planets in systems with multiple transiting planet candidates are significantly more likely to show TTVs than planets in stars with a single known transiting planet.  This supercedes the results of a similar analysis performed using just the first four months of Kepler data in Ford \etal (2011) that did not find a difference in the frequency of TTV signals between planet candidates transiting stars with one or multiple transiting planet candidates.  

Such comparisons provide information about the frequency of non-transiting planets with similar orbital distances as the transiting planets and can help break the degeneracy between planet abundance and the distributions of mutual inclinations (Lissauer \etal 2011b, Tremaine \& Dong 2011).  Since we do not know the distribution of non-transiting planets in these systems, we can not clearly identify whether this difference is due to observational bias.  For planetary systems with small, but non-zero mutual inclinations, the probability of multiple planets transiting is increased for closely-spaced systems that are also likely to yield large TTVs.  If such systems are common, then one would expect the rate of detectable TTVs to be greater when there are multiple transiting planets.  
On the other hand, the geometric probability of multiple planets transiting is drastically reduced for systems with modest mutual inclinations ($\ge5^\circ$; see figure 13 of  Lissauer et al. 2011b).  If the typical planetary system contained closely-spaced planets with mutual inclinations $\sim5^\circ$ or larger, then one would expect a similar rate of TTVs, regardless of the number of transiting planets, since the overwhelming majority of multiple planet systems would have only a single transiting planet.  Therefore, the measurement of an increased rate of TTVs for stars with multiple transiting planet candidates is important because it provides information about the distribution of mutual inclinations.  A complete analysis accounting for all relevant detection biases is important.  For example, the magnitude of TTVs can increase rapidly with increasing orbital eccentricity.  If there were a correlation between eccentricities and mutual inclinations, then the increased eccentricity of systems with large inclinations would increase the rate of detectable TTVs for multiple-planet systems with only one transiting planet.
A complete analysis accounting for all relevant detection biases will be quite involved and is beyond the scope of this paper.

The true frequency of TTVs is likely to rise for both samples with more detailed analyses and as the time span of \Kepler observations increases.  Further analysis will be required to understand how these results are affected by potential complications (e.g., some multiple planet systems might be overlooked due to TTVs interfering with their detection or transit time measurement).   

We divide the TTV data sets into two equal-sized samples (693) according to orbital period.  This results in dividing the sample at a period of 15.0d.  We find no indication of a difference in the frequency of planet candidates with some indication of potential TTVs (TTV Flag$\ge$1) among planet candidates with orbital period greater or less than 15.0d (12.1\% and 11.4\% for the short and long period sub-samples, respectively).  
If we restrict our attention to planet candidates with a more robust indication of potential TTVs (TTV Flag$\ge$2), then we find a $\sim$60\% greater rate of TTV candidates for the long-period sample (6.1\% and 9.7\% for the short and long period sub-samples, respectively).  Based on Monte Carlo simulations, the probability of these two subsets of KOIs have the same occurrence rate of TTV signals is $\lesssim$0.1\%.  
However, we find  no significant difference for the rate of long-term TTV trends among planet candidates with orbital period greater or less than 15.0d (i.e., 2.6\% and 3.0\% for the short and long period sub-samples, respectively).  It would not be unexpected for planet candidates with larger orbital periods to have a greater frequency of detectable TTVs, since the TTV amplitude typically scales with the orbital period.  However, our results do not provide strong support for such a trend.  

We divide the TTV data sets into two equal-sized samples according to photometric S/N of the transits.  This results in dividing the sample at a S/N of individual transits of 6.6.  We find $\sim22\%$ more planet candidates with some indication of potential TTVs (TTV Flag$\ge$1) among the high S/N sample and $\sim73\%$ more planet candidates with a more robust indication of potential TTVs (TTV Flag$\ge$2). We find a $\sim37\%$ larger rate of planet candidates with long-term TTV trends among the high S/N subsample.  The increased frequency of planet candidates with excess TTV scatter or long-term TTV trends is likely due to the higher quality transit timing precission for planet candidates with high S/N per transit.

We recommend that observers planning follow-up observations of \Kepler planet candidates check whether there are indications of significant TTVs.  While most KOIs show TTVs that are small compared to the transit duration, some systems (e.g., KOI 806) have TTVs so large that they can affect the scheduling of follow-up observations (e.g., Tingley \etal 2011; Fabrycky \etal 2012).  Based on synthetic planet populations (Lissauer \etal 2011b), we anticipate that $\sim15-26\%$ of planets could have TTVs as large as the transit duration.  Of course, the higher-order ephemerides should not be trusted to predict future transit times accurately, as extrapolations are often unreliable.  
 
On sufficiently long timescales, most TTV are likely better modeled with periodic functions.  Indeed, Steffen \etal (2012) have performed a complementary analysis using a sinusoidal model.  A periodic model is likely superior for TTV signals once they have been observed for multiple TTV cycles.  We caution TTV signatures can be quite complex (e.g., Veras \etal 2011), so one can not presume that a sinusoidal model is necessarily a better model until the TTV signal has been observed for multiple cycles.  Further, polynomial models require fewer model parameters and do not require searching over many possible TTV frequencies, making polynomial models more sensitive to low-amplitude, long-term TTV signals.  Thus, polynomial and sinusoidal TTV models are both important and complementary tools to search for TTV candidates.  We anticipate that the sinusoidal model is likely more appropriate for TTV candidates that require a $N_d\ge4$ degree model.  

The largest planetary TTV signatures are likely to arise in systems in or near mean motion resonances, but the associated timescales are typically multiple years (Agol \etal 2005; Holman \& Murray 2005; Veras \etal 2011; Veras \& Ford 2011).  Over timescales short compared to the TTV timescale, polynomial models can often approximate the dominant TTV signal with a minimal number of model parameters.  On the other hand, a periodic TTV model requires at least five model parameters (and a minimum of six transits to assess a model).  Searching over many possible TTV frequencies requires performing many statistical tests, reducing the sensitivity to small amplitude signals when only a single TTV cycle has been observed.  For planets with orbital periods approaching a year, it will be important to analyze TTV signals using as few transits as possible.  
Borucki \etal (2012) obtained early TTV constraints for a planet in the habitable zone using just 3 transit times.  
For tightly-packed systems with multiple transiting planets, a full TTV analysis may be possible even with only $\sim$6 transits per planet (e.g., Holman \etal 2010; Fabrycky \etal 2012).  Thus, we anticipate that polynomial TTV models will remain an important tool for assessing TTV candidates, particularly for planet candidates with long orbital periods, such as those approaching the habitable zone of solar-type stars.  

We anticipate that many of the planet candidates with long-term TTV trends (see  Fig.\ 1, 2 \& 3) are likely to be confirmed with additional analysis.  Such large and coherent trends are unlikely to be due to astrophysical noise or non-Gaussian measurement errors.  In some cases, further observations and/or TTV analysis may lead to the detection of additional non-transiting planets (e.g., Ballard \etal 2011), or even small planets with large TTVs that were not detected by the standard \Kepler planet search pipeline.  In other cases, long-term TTV trends may be due to perturbations from stellar or brown dwarf companions rather than additional planets.  It is also possible that a transiting planet candidate with TTVs is actually an eclipsing binary that is significantly diluted, so the transit depth is consistent with planetary transit, and a member of a triple star system (e.g., KOI 928.01; Steffen \etal 2011).  
It is less clear what fraction of the planet candidates with excess TTV scatter have dynamically induced TTVs, as opposed to an apparent TTV signal due to astrophysical noise (e.g., D{\'e}sert \etal 2011) or non-Gaussian measurement errors.  Nevertheless, we recommend that these candidates be subjected to more detailed TTV analysis, as some of these are planets in closely-packed systems with complex TTVs (e.g., Kepler-11; Lissauer \etal 2011a).  Such systems are particularly favorable for measuring planet masses and bulk density, so as to constrain the planet's composition and nature (i.e., distinguish between rocky versus gaseous planets).  

\acknowledgements  Funding for this mission is provided by NASA's Science Mission Directorate.  We thank the entire \Kepler team for the many years of work that is proving so successful.  
E.B.F acknowledges support by the National Aeronautics and Space Administration under grant NNX08AR04G issued through the \Kepler Participating Scientist Program.  
{\it Facilities:} \facility{Kepler}.

\newpage

\begin{deluxetable}{llll}
\tabletypesize{\scriptsize}
\tablecaption{Kepler TTV Metrics Catalog Format}
\tablewidth{0pt}
\tablehead{
\colhead{Column}   & \colhead{Format}    & \colhead{Description} }
\startdata
 1 & KOI       & F7.2 & Kepler Object of Interest Identifier \\ 
 2 & S/N       & F7.1 & Photometric signal-to-noise for typical transit \\ 
 3 & nTT       & I5 & Number of transits used for analysis \\ 
 4 & $T_{dur}$ & F5.1  & Estimated transit duration (Batalha et al.\ 2012) \\ 
 5 & $N_d$     & I2 & Degree of polynomial selected for ephemeris \\ 
 6 & TTV Flag  & I4 & TTV Flag\tablenotemark{a}\\ 
 7 & $E_{\mathrm{lin}}$         & F12.6 & Epoch of best-fit linear ephemeris \\ 
 8 & $\sigma_{E_{\mathrm{lin}}}$ & F9.6 & Uncertainty in $E_{\mathrm lin}$ \\ 
 9 & $P_{\mathrm{lin}}$         & F11.6 & Period of best-fit linear ephemeris \\ 
10 & $\sigma_{P_{\mathrm{lin}}}$ & F9.6 & Uncertainty in $P_{\mathrm lin}$ \\ 
11 & $E_{N_d}$          & F12.6 & Epoch of best-fit $N_d$-degree polynomial ephemeris  \\ 
12 & $\sigma_{E_{N_d}}$  & F9.6 & Uncertainty in $E_{N_d}$ \\ 
13 & $P_{N_d}$          & F11.6 & Period of best-fit $N_d$-degree polynomial ephemeris \\ 
14 & $\sigma_{P_{N_d}}$ & F9.6 & Uncertainty in $P_{N_d}$ \\ 
15 & $c_{2}$            & 1X,E10.2 & Coefficient of $P_{N_d} \times n^2$ for best-fit $N_d$-degree polynomial ephemeris\tablenotemark{b} \\ 
16 & $\sigma_{c_{2}}$    & 1X,E10.2 & Uncertainty in $c_{2}$\tablenotemark{b} \\ 
17 & $c_{3}$            & 1X,E10.2 & Coefficient of $P_{N_d} \times n^3$ for best-fit $N_d$-degree polynomial ephemeris\tablenotemark{c} \\ 
18 & $\sigma_{c_{3}}$    & 1X,E10.2 & Uncertainty in $c_{3}$\tablenotemark{c} \\ 
19 & $c_{4}$            & 1X,E10.2 & Coefficient of $P_{N_d} \times n^4$ for best-fit $N_d$-degree polynomial ephemeris\tablenotemark{d}\\ 
20 & $\sigma_{c_{4}}$    & 1X,E10.2 & Uncertainty in $c_{4}$\tablenotemark{d} \\ 
21 & $\sigma_{TT}$      & F6.1 & Median transit time uncertainty in minutes \\
22 & MAD                & F6.1 & Median absolute deviation of transit times from a linear ephemeris in minutes\\
23 & WRMS               & F6.1 & Weighted root mean square deviation of transit times from a linear ephemeris in minutes \\
24 & $p_{\chi^2,{\mathrm{lin}}}$   & 1X,E10.2 & $p$-value for a $\chi^2$-test relative to linear ephemeris \\ 
25 & $p_{\chi'^2,{\mathrm{lin}}}$  & 1X,E10.2 & $p$-value for a $\chi^2$-test using $\chi'^2$ relative to linear ephemeris\tablenotemark{e} \\ 
26 & $p_{\chi^2,{\mathrm{quad}}}$  & 1X,E10.2 & $p$-value for a $\chi^2$-test relative to quadratic ephemeris \\ 
27 & $p_{\chi'^2,{\mathrm{quad}}}$ & 1X,E10.2 & $p$-value for a $\chi^2$-test using $\chi'^2$ relative to quadratic ephemeris\tablenotemark{e} \\ 
28 & $p_{\chi^2,{\mathrm{cube}}}$  & 1X,E10.2 & $p$-value for a $\chi^2$-test relative to cubic ephemeris \\ 
29 & $p_{\chi'^2,{\mathrm{cube}}}$ & 1X,E10.2 & $p$-value for a $\chi^2$-test using $\chi'^2$ relative to cubic ephemeris\tablenotemark{e} \\ 
30 & $p_{\chi^2,{\mathrm{quar}}}$  & 1X,E10.2 & $p$-value for a $\chi^2$-test relative to quartic ephemeris \\ 
31 & $p_{\chi'^2,{\mathrm{quar}}}$ & 1X,E10.2 & $p$-value for a $\chi^2$-test using $\chi'^2$ relative to quartic ephemeris\tablenotemark{e} \\ 
32 & $p_{F,{\mathrm{quad,lin}}}$  & 1X,E10.2 & $p$-value for an $F$-test comparing linear and quadratic ephemerides\tablenotemark{f} \\ 
33 & $p_{F,{\mathrm{cube,lin}}}$  & 1X,E10.2 & $p$-value for an $F$-test comparing linear and cubic ephemerides\tablenotemark{f} \\ 
34 & $p_{F,{\mathrm{cube,quad}}}$ & 1X,E10.2 & $p$-value for an $F$-test comparing quadratic and cubic ephemerides\tablenotemark{f} \\ 
35 & $p_{F,{\mathrm{quar,lin}}}$ & 1X,E10.2 & $p$-value for an $F$-test comparing linear and quartic ephemerides\tablenotemark{f} \\ 
36 & $p_{F,{\mathrm{quar,quad}}}$ & 1X,E10.2 & $p$-value for an $F$-test comparing quadratic and quartic ephemerides\tablenotemark{f} \\ 
37 & $p_{F,{\mathrm{quar,cube}}}$ & 1X,E10.2 & $p$-value for an $F$-test comparing cubic and quartic ephemerides\tablenotemark{f} \\ 
\enddata
\tablenotetext{a}{Sum of following flags:
   1 if $p_{\chi^2}<10^{-3}$,
   2 if $p_{\chi'^2}<10^{-3}$,
   4 if $MAD \ge 5 \sigma_{TT}$,
   8 if $p_{F,{\mathrm{lin,quad}}} < 0.02$ AND nTT$\ge4$,
  16 if $p_{F,{\mathrm{lin,cube}}} < 0.02$ AND nTT$\ge5$ AND IF(8-bit flag is set, $p_{F,{\mathrm{quad,cube}}} < 0.02$, TRUE), and
  32 if $p_{F,{\mathrm{lin,quar}}} < 0.02$ AND nTT$\ge8$ AND IF(16-bit flag is set, $p_{F,{\mathrm{cube,quar}}} < 0.02$, IF(8-bit flag is set, $p_{F,{\mathrm{quad,quar}}} < 0.02$, TRUE)).}
\tablenotetext{b}{Zero if $N_d<$2}
\tablenotetext{c}{Zero if $N_d<$3}
\tablenotetext{d}{Zero if $N_d<$4}
\tablenotetext{e}{For description of $\chi'^2$ statistic and tests, see \S~3.1 of Ford \etal 2011}
\tablenotetext{f}{Value of -1.0 if insufficient number of transits}
\tablecomments{All ephemerides and TTV statistics are calculated from \Kepler observations up to and including quarter six (Rowe \etal 2012).} 
\label{tabKey}
\end{deluxetable}

\newpage

\begin{figure*}
\plotone{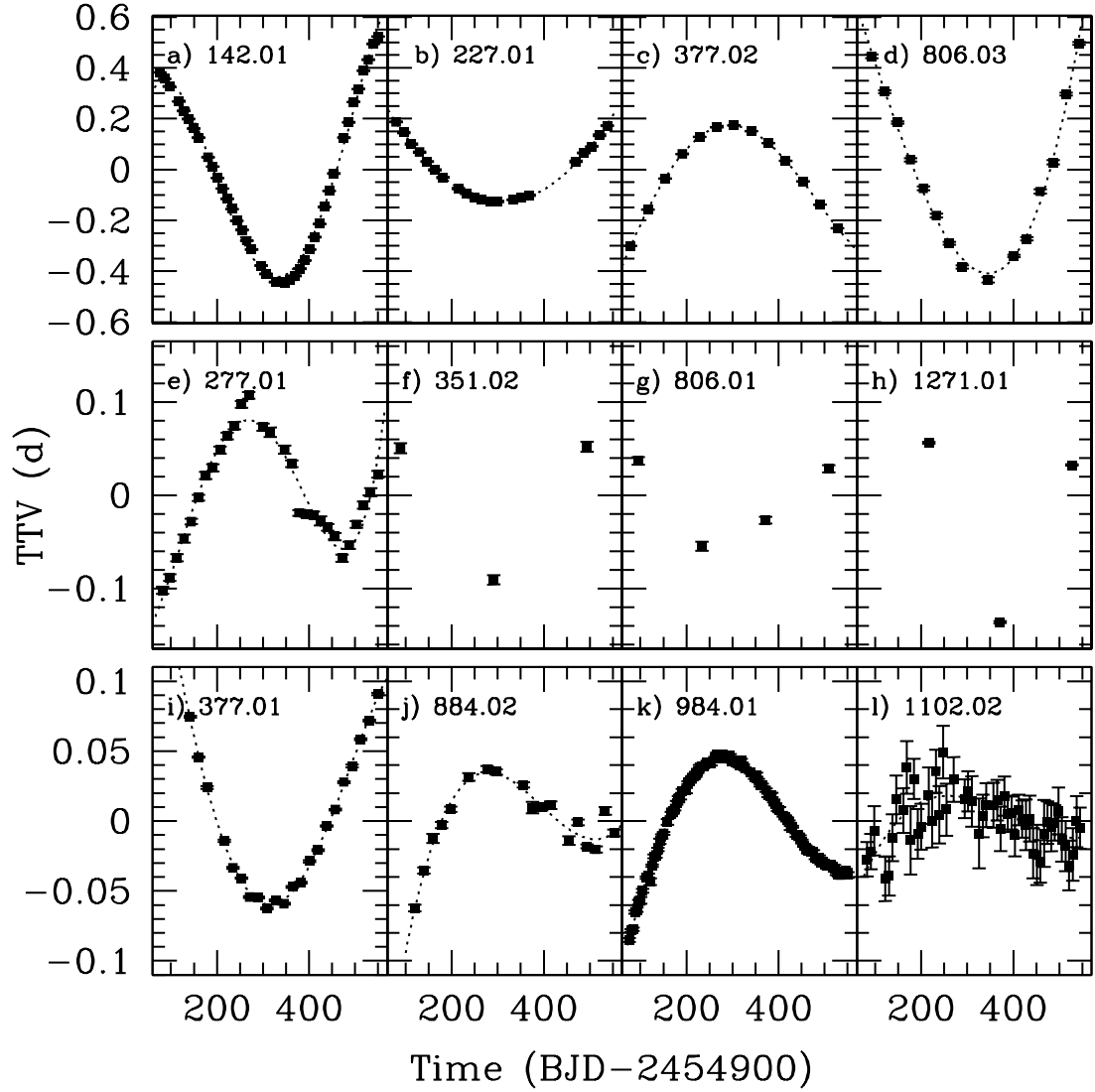}
\caption{Transit times for planet candidates with long-term TTV trends.  Kepler-9c (panel c), Kepler-30b (panel d), Kepler-30d (panel g), Kepler-9b (panel i) and Kepler-24b (panel l)  have already been confirmed via TTVs (Holman \etal 2010; Fabrycky et al. 2012; Ford et al.\ 2012).  
Dotted lines show the best-fit polynomial model given in Table \ref{tabKey}.  For KOIs 351.01 (panel f), 806.01 (panel g) and 1271.01 (panel h) there is no polynomial model, since a polynomial model (of degree 4 or less) did not significantly improve upon the linear ephemeris.  Similar figures for additional planet candidates will be available at \hbox{http://www.astro.ufl.edu/$\sim$eford/data/kepler/}.
}
\label{fig1}
\end{figure*}

\begin{figure*}
\plotone{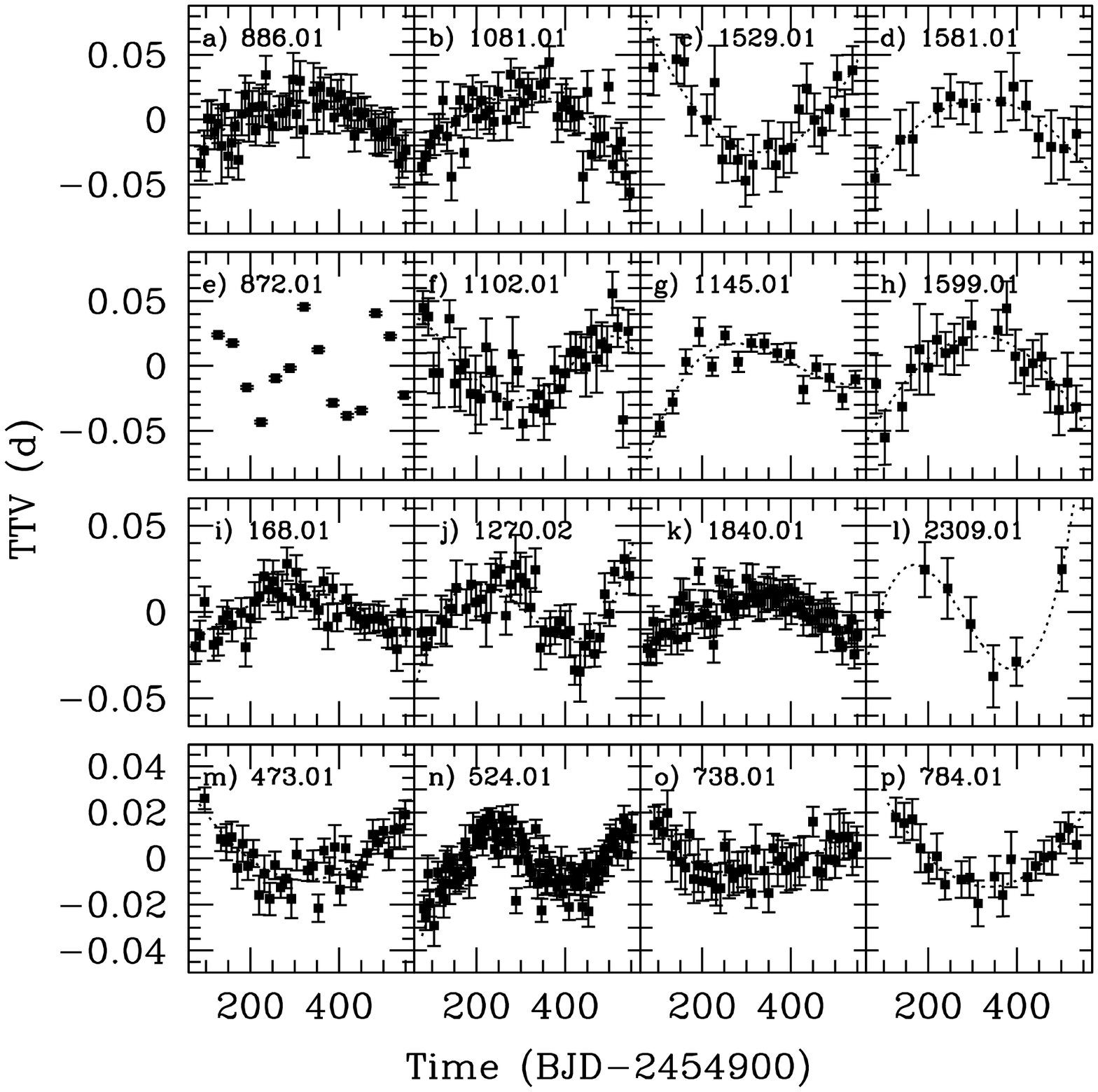}
\caption{Transit times for planet candidates with long-term TTV trends.  KOI 872b (panel e), Kepler-24c (panel f), Kepler-23c (panel i) and Kepler-29b (panel o) have already been confirmed via TTVs (Nesvorny et al.\ 2012; Ford et al.\ 2012; Fabrycky et al.\ 2012).  
Dotted lines show the best-fit polynomial model given in Table \ref{tabKey}.  For KOI 872.01 (panel e) there is no polynomial model, since a polynomial model (of degree 4 or less) did not significantly improve upon the linear ephemeris.  
}
\label{fig2}
\end{figure*}

\begin{figure*}
\plotone{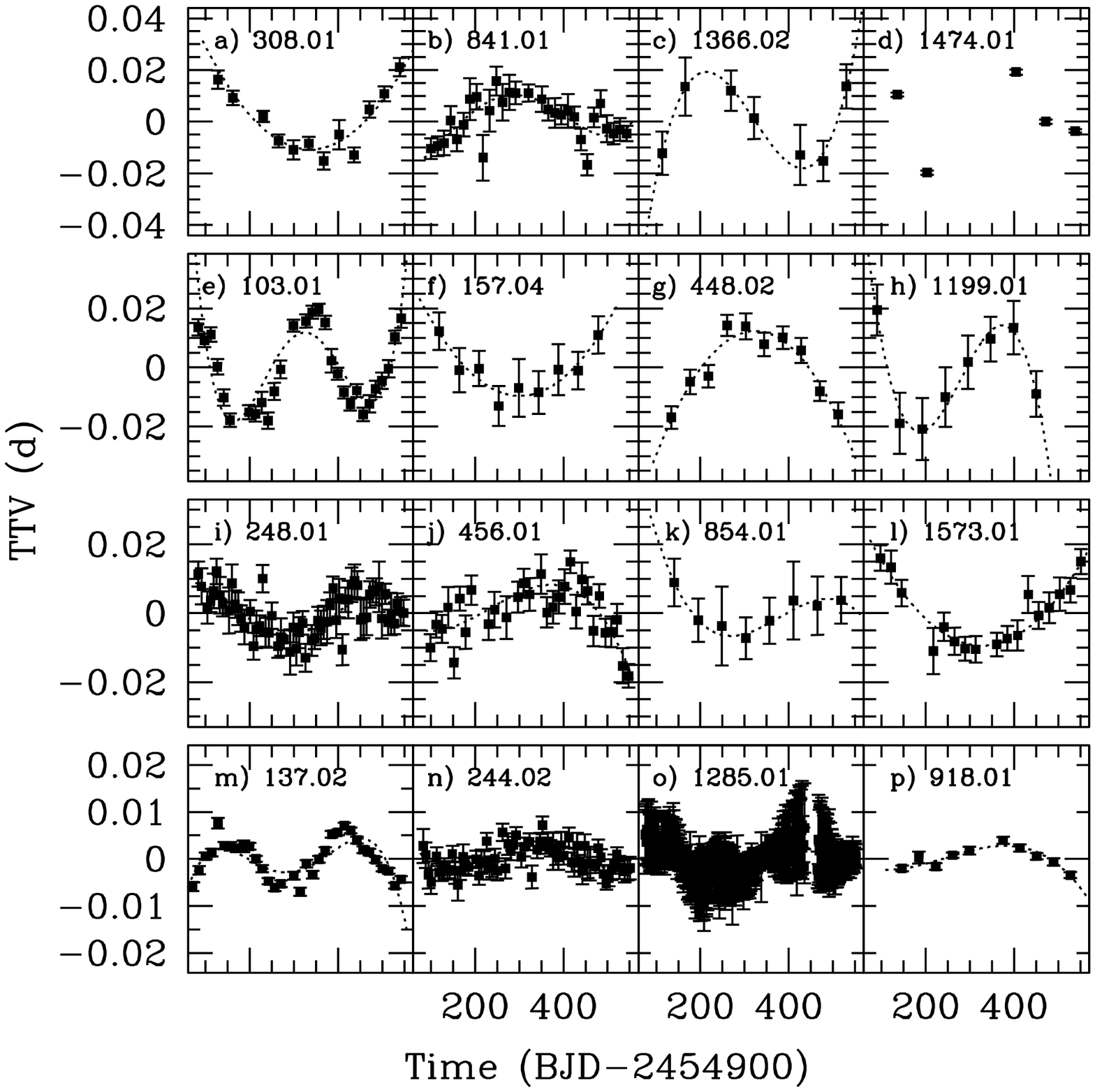}
\caption{Transit times for planet candidates with long-term TTV trends.  Kepler-27b (panel b), Kepler-11f (panel f), Kepler-18d (panel m) and Kepler-25b (panel n) have already been confirmed via TTVs (Steffen et al.\ 2012; Lissauer et al.\ 2011a; Cochran \etal 2011).  
Dotted lines show the best-fit polynomial model given in Table \ref{tabKey}.  For KOI 1474.01 (panel d) there is no polynomial model, since a polynomial model (of degree 4 or less) did not significantly improve upon the linear ephemeris.  
}
\label{fig3}
\end{figure*}


\end{document}